\documentclass[showpacs,preprint,aps]{revtex4}
\usepackage{mathrsfs}
\usepackage{amsmath}
\usepackage{stmaryrd}
\usepackage{amssymb}
\usepackage[dvips]{graphicx}

\begin{document}
\title{Nonlinear theory of intense laser-plasma interactions modified
 by vacuum-polarization effects}
\author{Wenbo~Chen$^{1)}$}
\email{cwb@shu.edu.cn}
\author{Zhigang~Bu$^{1)}$}
\author{Hehe~Li$^{1)}$}
\author{Yuee~Luo$^{1)}$}
\author{Peiyong~Ji$^{1)2)}$}
\email{pyji@staff.shu.edu.cn}
\affiliation{1) Department of
Physics, Shanghai University, Shanghai 200444, China\\
2) The Shanghai Key Lab of Astrophysics, Shanghai  200234, China}

\begin{abstract}
The classical nonlinear laser-plasma interaction theory is corrected. Given the effects of vacuum polarization (induced by extreme laser) as nonlinear media response, one-dimensional wave equations of a monochromatic laser field are derived from the Heisenberg-Euler Lagrangian density and a derivative correction with the first order quantum electrodynamic (QED) effects. A more suitable model to formulate the interactions of extreme laser and high-energy-density plasma is developed. In the results, the enhanced effect of vacuum polarization will be discussed and shown.
\end{abstract}

\maketitle

\setlength\arraycolsep{2pt}
\section{Introduction}

With the growing advancement of laser technology, the power of the next-generation optical laser systems will reach about $10\sim100PW$ levels, and after focusing to $1 \mu m^{2}$ the intensity of the laser beam is expected to exceed $10^{25}W/cm^{2}$\cite{3}. Meanwhile, with the electron densities of the plasma approaching unprecedented $10^{26}cm^{-3}$ achieved under the present experimental conditions, X-rays with high energies of $E \geq 10keV$ are required to penetrate through the plasma\cite{4}. Taken into account the effects of quantum vacuum in the interactions of extreme laser with high-energy-density plasma, the classic theory should be modified accordingly.

As fundamental concepts in quantum field theory, the effects of quantum vacuum were neglected for many years but are now beginning to be hot issues for the improvement of experiment techniques. The effective Lagrangian density for quantum vacuum corrections in electromagnetic field was presented by Heisenberg and Euler as early as 1936\cite{9}, and improved by Schwinger in 1951\cite{10}. Significantly observable phenomena of quantum vacuum such as vacuum breakdown, electron-positron pairs created from vacuum, need to impose a super-strong background field. It is the Schwinger critical field $E_{Sc}=m_{e}^{2}c^{3}/{e\hbar}\approx 10^{16}V/cm$, corresponding to the laser intensity $I_{cr}=E_{cr}^{2}/8\pi=4.6\times10^{29}W/cm^{2}$. However, for the present experimental conditions it is still very difficult to achieve the critical field-strength, and the realization of quantum vacuum lacks experimental validations. Recent years, many researchers tried to put forward reasonable models and observable effects which focus on the effects of vacuum polarization induced by intense laser pulses\cite{11, 12}. The vacuum-polarization effects, requiring laser intensity much lower than the critical value $I_{cr}$, are expected to be observed in the near future, and the predictions of QED theory on quantum-vacuum phenomena will be verified. In all the models about the effects of vacuum-polarization, the interaction of intense laser and plasma is an effective issue. Such research exhibiting the effects of QED vacuum via plasma characteristics has indeed made some progress in recent years\cite{14, 15, 16}.

When we investigate the processes of intense laser-plasma interaction, an important parameter of the laser intensity $a_{0}$ is particularly noteworthy. Defined as the peak amplitude of the laser field, normalized vector potential $\mathbf{a}=e\mathbf{A}/m_{e}c^{2}$, the parameter $a_{0}$ and the peak laser light intensity $I_{0}$ possess the relationship of $a_{0}^{2}\sim 7.3\times 10^{-19}[\lambda(\mu m)]^{2}I_{0}(W/cm^{2})$. When $a_{0}\ll 1$, a linear theory is only required for such interaction\cite{5,6}. Once in the regime of $a_{0}\geq 1$, the electronic behaviors are relativistic and the intense laser-plasma interaction must be described by a nonlinear theory. A valid one-dimensional analytical model was presented in the 1990s by Esarey et al.\cite{7}. In the case of three-dimensional, some results via numerical simulation were also obtained\cite{8} and subsequent issues related to nonlinear laser-plasma interaction, e.g., wave breaking, electron acceleration, multiple laser beam and a series of problems were discussed. If the laser light intensity is further raised to make $a_{0}^{2}\gg 1$, new phenomena for the interaction of laser (into the regime of extreme light sources) with plasma cannot be explained by the classical nonlinear theory. So it is necessary to complement and extend earlier works.

The effects of quantum vacuum induced by the laser with extreme intensity and high frequency can not be neglected. Proper corrections due to the vacuum-polarization effects are considered in the coupling equations for describing the interactions of extreme laser and high-energy-density plasma. In this paper, the vacuum-polarization effects, similar to the response in media, are introduced via Heisenberg-Euler (H-E) correction term and derivative correction term in the Lagrangian density of electromagnetic fields. A 1D nonlinear coupling model and the corresponding results are obtained for a more accurate description. The characteristics of these two kinds of corrections are compared, and the results are discussed in detail with the parameters of the most likely areas, such as inertial confinement fusion (ICF) as well as the compact object as dwarf star, for the valid observation in high-energy-density plasma physics. Compared with the classical theory of 1D nonlinear laser-plasma interaction\cite{7}, our model with the contribution of vacuum-polarization effects is more effective approach to describe the practical situation of extreme light and high-energy-density plasma.

The paper is organized as follows. In Sec. II, introducing the correction terms of vacuum polarization into the equations of electromagnetic field, a 1D wave equations of a monochromatic paraxial laser field is derived. In Sec. III, a nonlinear coupling model to formulate the intense laser-plasma interaction under the influence of vacuum-polarization effects is obtained. Finally in Sec. IV, the influence of the vacuum-polarization corrections is investigated, and the effective way to enhance the effects in high-energy-density physics is explored.

\section{Basic Equations of Laser Field with QED Vacuum Effects}

A set of wave equations for laser field has been derived to formulate the extreme light behavior. In such wave equations, two kinds of QED vacuum effects have been taken into account. (i) The high intensity effect of vacuum polarization is formulated by H-E Lagrangian density when the laser intensity is extreme but still much lower than Schwinger critical intensity, and (ii) the effect of rapid field variations can be illustrated by the derivative correction Lagrangian density when the laser frequency is ultra-high but still much lower than Compton frequency $\omega_{e}=mc^{2}/ \hbar$. For all these cases, the Lagrangian density of electromagnetic fields can be represented as follows\cite{14},
\begin{eqnarray}
\mathcal{L} &=& \mathcal{L}_{0}+\delta \mathcal{L}_{HE}+\delta
            \mathcal{L}_{D}-\frac{1}{c} j^{\mu}A_{\mu}, \nonumber\\
            &=& -\displaystyle\frac{1}{16\pi}F_{\mu\nu}F^{\mu\nu}+
            \displaystyle\frac{\kappa}{128\pi}\left[4\left(F_{\mu\nu}
            F^{\mu\nu}\right)^{2}+7\left(F_{\mu\nu}\hat{F}^{\mu\nu}
            \right)^{2}\right]\nonumber\\
            &+&\displaystyle\frac{\sigma}{8\pi}\left[\left(\partial_
            {\mu}F^{\mu\nu}\right)\left(\partial_{\tau}F^{\tau}_{~\nu}
            \right)-F_{\mu\nu} \Box F^{\mu\nu}\right]-\frac{1}{c}
            j^{\mu}A_{\mu},
            \label{eq:1}
\end{eqnarray}
where $\mathcal{L}_{0}$ is the classical Lagrangian density, $\delta \mathcal{L}_{HE}$ is H-E correction (composed of two Lorentz invariants, $F_{\mu\nu}F^{\mu\nu} = -2(\mathbf{E}^{2} -\mathbf{B}^{2})\equiv -4 \mathcal{F}$ and $\ F_{\mu\nu}\hat{F}^{\mu\nu}= -4\mathbf{E}\cdot\mathbf{B}\equiv -4 \mathcal{G}$), and $\delta \mathcal{L}_{D}$ is derivative correction. $F^{\mu\nu}=\partial^{\mu}A^{\nu}-\partial^{\nu}A^{\mu}$ is the four-dimensional electromagnetic tensor, $\hat{F}^{\mu\nu}=\epsilon^{\mu\nu\lambda\tau} F_{\lambda\tau}/{2}, \Box=\partial_{\alpha}\partial^{\alpha}$, and $j^{\mu}$ is four-dimensional current density, where $A^{\mu}$ is contravariant electromagnetic potential. The parameter
$\kappa={\alpha^{2}\hbar^{3}}/{45\pi m_{e}^{~4}c^{5}}$ gives the H-E coupling, the parameter $\sigma={4\alpha\hbar^{2}}/{15\pi m_{e}^{~2}c^{5}}$ gives the derivative effects of vacuum
polarization, and $\alpha=e^{2}/\hbar c \approx{1}/{137}$ is the fine structure constant.

From the Euler-Lagrange equation, ${\partial \mathcal{L}}/{\partial A_{\mu}}-\partial_{\nu}[{\partial \mathcal{L}}/{\partial(\partial_{\nu}A_{\mu})}]=0$, the field equations with vacuum-polarization corrections can be obtained as
\begin{equation}
    \left(1+\sigma\Box\right)\partial_{\nu}F^{\nu\mu}+\kappa\partial_{\nu}
    \left(4\mathcal{F}F^{\nu\mu}+7\mathcal{G}\hat{F}^{\nu\mu}\right)
    =\frac{4\pi}{c}j^{\mu}
    \label{eq:2},
\end{equation}
i.e.,
\begin{eqnarray}
    \left[1+\sigma\left(\nabla^{2}-\displaystyle\frac{1}{c^{2}}\partial_{t}^{~2}
    \right)\right]\nabla\cdot\mathbf{E} & = & -4\pi e\delta n -4\pi \rho_{vac},
    \label{eqs:1-1}\\
    \left[1+\sigma\left(\nabla^{2}-\displaystyle\frac{1}{c^{2}}\partial_{t}^{~2}
    \right)\right]\left(-\displaystyle\frac{1}{c}\partial_{t}\mathbf{E}+\nabla\times
    \mathbf{B}\right) & = & -\frac{4\pi}{c}en\mathbf{v}+\frac{4\pi}{c}\mathbf{j}_{vac},
    \label{eqs:1-2}
\end{eqnarray}
where $\delta n=n-n_{0}$ is the density perturbation of electron, $n_{0}$ is the ambient plasma density, and $\mathbf{v}$ is the electron fluid velocity. Due to the high intensity effect of vacuum
polarization, the effective vacuum ``sources'' (vacuum ``current'' $\rho_{vac}$ and vacuum ``charge'' $\mathbf{j}_{vac}$) appearing in Eqs.\eqref{eqs:1-1} and \eqref{eqs:1-2} can be written as
\begin{eqnarray}
    \rho_{vac} & = & \nabla\cdot\frac{\kappa}{4\pi}(4\mathcal{F}\mathbf{E}+
    7\mathcal{G}\mathbf{B}),
    \label{eqs:2-1}\\
    \mathbf{j}_{vac} & = & \partial_{t}\frac{\kappa}{4\pi}(4\mathcal{F}\mathbf{E}
    +7\mathcal{G}\mathbf{B})-c\nabla\times\frac{\kappa}{4\pi}(4\mathcal{F}\mathbf{B}
    -7\mathcal{G}\mathbf{E}).
    \label{eqs:2-2}
\end{eqnarray}

Assuming light beams to be monochromatic and paraxial, and neglecting self-focusing and diffraction effects in the beam transmission, the propagation of laser pulses can be characterized by 1D model, i.e., $\nabla \rightarrow \mathbf{e}_{z} \partial/\partial z=\mathbf{e}_{z}\partial_{z}$ (for light propagating along $z$ axis). Using the Coulomb gauge, $\nabla\cdot\mathbf{A}=0$, and considering the strong laser pulses with large amplitude, the longitudinal component (the one perpendicular to $z$ axis) of the vector potential is much larger than the transverse one, $|\mathbf{A}_{z}| \ll |\mathbf{A}_{\perp}|$, and then $\mathbf{A}\approx\mathbf{A}_{\perp}=A\mathbf{e}_{\perp}$. The effective
polarization $\mathbf{P}_{vac}$ and magnetization $\mathbf{M}_{vac}$ induced by vacuum ``sources'' are written simply as
\begin{eqnarray}
    \mathbf{P}_{vac} & = & -\frac{\kappa}{\pi c}\mathcal{F}\partial_{t}A\mathbf{e}_{\perp},
    \label{eqs:3-1}\\
    \mathbf{M}_{vac} & = &  \frac{\kappa c}{\pi}\mathcal{F}\partial_{z}A(\mathbf{e}_{z}\times\mathbf{e}_{\perp}).
    \label{eqs:3-2}
\end{eqnarray}

In the 1D model, the electromagnetic potentials of the laser beam are demonstrated as $A(z,t)=A_{0}(z,t)~e^{i\Psi}$ and $\Phi(z,t)=\Phi_{0}(z,t)~e^{i\Psi}$, where $\mathcal{A}_{0}(A_{0},\phi_{0})$ and $\Psi=-\omega t+kz+\psi(z,t)$ are the amplitude and the phase of the laser potential, respectively. The perturbations of the amplitude $\delta\mathcal{A}_{0}(\delta A_{0}, \delta \phi_{0})$ and the phase $\delta \psi_{0}$ induced by the vacuum-polarization effects are much small-varying, i.e., $|\delta \mathcal{A}_{0}| \ll |\mathcal{A}_{0}|$ and $|\delta\psi| \ll |\Psi|$. So the first-order differential terms of the amplitude and the phase will only be kept when we calculate the corrections of vacuum ``source'', and the approximate relations are derived as $\partial_{i}A\partial_{j}A \approx (\partial_{i}\partial_{j}A)A,\ \partial_{i}\mathcal{\phi}\partial_{j}\phi \approx (\partial_{i}\partial_{j} \phi)\phi$, and $\partial_{i}A\partial_{j}\phi \approx (\partial_{i}\partial_{j}A)  \phi \approx (\partial_{i}\partial_{j}\phi)A \approx \partial_{i}\phi \partial_{j}A$, where $i,j=t,z$.

Since the H-E coupling coefficient $\kappa$ is much small ($\kappa\sim 10^{-32}cm \cdot s^{2}/g$) and the laser field is much lower than $E_{Sc}$, the intensity contribution of vacuum-polarization corrections (vacuum ``source'') can be written as a perturbation term. Neglecting the higher-order terms ($\mathcal{O}(\kappa)$) in the equations, the vacuum ``source'' are simplified as follows, $\rho_{vac} \approx -6\kappa \cdot 4\pi e\left(\delta n\Phi-n\beta_{\perp}A\right)e\delta n$ and $\mathbf{j}_{vac} \approx 6\kappa \cdot 4\pi e\left(\delta n\Phi-n \beta_{\perp}A\right)en(v_{z}\mathbf{e}_{z}+v_{\perp}\mathbf{e}_{\perp})$, where $\delta n\sim \partial_{z}^{~2}\Phi, n{\beta_{\perp}}\sim\left(\partial_{z}^{~2}-{\partial_{t}^{~2}}/{c^{2}}\right)A$. Similarly, considering the coupling coefficient $\sigma$ of the derivative correction is small (about~$10^{-24}cm^{2}$) and the laser frequency is much lower than $\omega_{e}$, the first-order term of $\sigma$ is only kept in the equations. Finally from Eqs.\eqref{eqs:1-1} and \eqref{eqs:1-2}, the wave equations of a monochromatic paraxial light beam with vacuum-polarization corrections can be represented as
\begin{eqnarray}
  \begin{array}{rcl}
    \left(\partial_{z}^{~2}-\displaystyle\frac{1}{c^{2}}\partial_{t}^{~2}\right)a & = & k_{p}^{~2}\displaystyle\frac{n}{n_{0}}\beta_{\perp}-6\kappa (4\pi e)^{2}(\delta
    n\phi-n\beta_{\perp}a)n\beta_{\perp}-k_{p}^{~2}\sigma\left(\partial_{z}^{~2}-\displaystyle\frac{1}{c^{2}}\partial_{t}^{~2}\right)
    \left(\displaystyle\frac{n}{n_{0}}\beta_{\perp}\right),\\
    \displaystyle\frac{1}{c}\partial_{t}\partial_{z}\phi & = & -k_{p}^{~2}\displaystyle\frac{n}{n_{0}}\beta_{z}+6\kappa (4\pi e)^{2}(\delta
    n\phi-n\beta_{\perp}a)n\beta_{z}+k_{p}^{~2}\sigma\left(\partial_{z}^{~2}-\displaystyle\frac{1}{c^{2}}\partial_{t}^{~2}\right) \left(\displaystyle\frac{n}{n_{0}}\beta_{z}\right),\\
    \partial_{z}^{~2}\phi & = & k_{p}^{~2}\displaystyle\frac{\delta n}{n_{0}}-6\kappa (4\pi e)^{2}(\delta n\phi-n\beta_{\perp}a)\delta
    n-k_{p}^{~2}\sigma\left(\partial_{z}^{~2}-\displaystyle\frac{1}{c^{2}}\partial_{t}^{~2}\right) \left(\displaystyle\frac{\delta n}{n_{0}}\right),
    \label{eqs:5}
  \end{array}\nonumber\\
\end{eqnarray}
where $\phi= \left({e}/{m_{e}c^{2}}\right)\Phi$ and $a= \left({e}/{m_{e}c^{2}}\right)A$ are the normalized electromagnetic potentials, $k_{p}^{~2}={\omega_{p}^{~2}}/{c^{2}}={4\pi
e^{2}n_{0}}/{m_{e}c^{2}}$, $\beta_{\perp}={v_{\perp}}/{c}$, and $\beta_{z}={v_{z}}/{c}$.

\section{The Response of Vacuum Polarization in the Nonlinear Intense Laser-Plasma Interactions}

In the regime of extreme lasers, the processes of intense laser-plasma interactions are something different from the classical theories. Especially when the intensity of laser beams is approximated to the Schwinger critical intensity, the QED vacuum effects will play a significant role. In this section, considering the proper corrections of vacuum polarization, a 1D nonlinear coupling model is found to describe the self-consistent interactions of intense laser with plasma .

To represent the plasma wave (laser wakefield) generated by this interaction, the fluid plasma model is generally used\cite{17}. The perturbation of the plasma electrons (plasma wave) generated by the laser fields obeys the following hydrodynamic equations,
\begin{eqnarray*}
    \partial_{t}n+\nabla\cdot(n\mathbf{v}) & = & 0,\\
    \partial_{t}\mathbf{p}+(\mathbf{v}\cdot\nabla)\mathbf{p} & = & -e\left(\mathbf{E}+\frac{\mathbf{v}}{c}\times\mathbf{B}\right),
\end{eqnarray*}
where $\mathbf{p}=m\mathbf{v}\gamma$ is the momentum of electron, $\gamma=\left(1-\beta^{2}\right)^{-{1}/{2}}$ is the relativistic factor and $\beta={v}/{c}$ .

Taking account of the vacuum-polarization effects, a driving laser pulse with extreme intensity can be formulated by Eq.\eqref{eqs:5}. Assuming that the laser wakefield is under the conditions of paraxial approximation, the velocity of the electrons in wakefield can be decomposed into transverse and longitudinal, $\beta\mathbf{e}=\beta_{\perp}\mathbf{e}_{\perp}+\beta_{z}\mathbf{e}_{z}$, in 1D model.  So the continuity equation is
\begin{equation}
    \frac{1}{c}\partial_{t}n+\partial_{z}(n\beta_{z}) = 0, \label{eq:3}\\
\end{equation}
and the momentum equation of electronic fluid in laser wakefield is divided into two components (transverse and longitudinal),
\begin{equation}
    \frac{1}{c}\partial_{t}(\gamma\beta_{\perp})+\beta_{z}\partial_{z}(\gamma\beta_{\perp}) = \frac{1}{c}\partial_{t}a+\beta_{z}(\partial_{z}a), \label{eqs:6-1}
\end{equation}
and
\begin{equation}
    \frac{1}{c}\partial_{t}(\gamma\beta_{z})+\beta_{z}\partial_{z}(\gamma\beta_{z}) = \partial_{z}\phi-\beta_{\perp}(\partial_{z}{a}). \label{eqs:6-2}
\end{equation}

The transverse component, perpendicular to the direction of wake-field propagation, represents the rapid quiver motion of plasma electrons under the disturbances of intense laser pulse. From Eq.\eqref{eqs:6-1}, normalized vector potential $a=\gamma\beta_{\perp}$ and relativistic factor $\gamma^{2}=1-\beta_{z}^{~2}-\beta_{\perp}^{~2}=\left(1+a^{2}\right)/\left(1-\beta_{z}^{~2}\right)$ can be easily obtained.

The longitudinal component mainly formulates the drift motion of electrons, the acceleration or deceleration motion in the plasma wakefield, induced by the intense laser along its
propagation direction. The vacuum-polarization effects fluctuate the plasma wave by influencing the driving laser directly.

When the intensity of driving laser beam exceeds a certain value (about $10^{18}W/cm^{2}$ for $1\mu m$ wave length), i.e., $a \sim 1$, the electron motion in the laser wakefield is relativistic. Further, when $a \gg 1$, the longitudinal drift velocity of electrons (approaching the speed of light) is much larger than the transverse quiver velocity, namely, $\beta_{\perp} \ll \beta_{z} \sim 1$. From Eq.\eqref{eqs:6-2}, the fluid equations can be obtained in the form
\begin{eqnarray}
    \frac{1}{c}d_{t}\beta_{z} & = & -\frac{1}{2\gamma^{2}}\left(\partial_{z}+\frac{1}{c}\beta_{z}\partial_{z}\right)a^{2}+\frac{1}{\gamma}\left(1-\beta_{z}^{~2}\right)\partial_{z}\phi,
    \label{eqs:7-1}\\
    \frac{1}{c}d_{t}\gamma & = & \beta_{z}\partial_{z}\phi+\frac{1}{2c\gamma}\partial_{z}a^{2}, \label{eqs:7-2}
\end{eqnarray}
where $d_{t}=\partial_{t}+c\beta_{z}\partial_{z}$. It is convenient to perform a transformation from the laboratory frame to the speed of light frame, $\xi=z-ct,\ \tau=t$. The potential wave Eq.\eqref{eqs:5} of the intense laser beam can be rewritten as
\begin{eqnarray}
  \begin{array}{rcl}
    \displaystyle\left(\frac{2}{c}\partial_{\xi}\partial_{\tau}-\frac{1}{c^{2}}\partial_{\tau}^{~2}\right)a & = & k_{p}^{~2}\displaystyle\frac{n}{n_{0}\gamma}a-6\kappa (4\pi
    e)^{2}\left(\delta n\phi-\frac{n}{\gamma}a^{2}\right)\frac{n}{\gamma}a-k_{p}^{~2}\sigma\left(\frac{2}{c}\partial_{\xi}\partial_{\tau}-\frac{1}{c^{2}}\partial_{\tau}^{~2}\right)
    \displaystyle\frac{n}{n_{0}\gamma}a,\\
    \displaystyle\left(\frac{1}{c}\partial_{\xi}\partial_{\tau}-\frac{1}{c^{2}}\partial_{\xi}^{~2}\right)\phi & = & -k_{p}^{~2}\displaystyle\frac{n}{n_{0}}\beta_{z}+6\kappa (4\pi
    e)^{2}\left(\delta n\phi-\frac{n}{\gamma}a^{2}\right)n\beta_{z}+k_{p}^{~2}\sigma\left(\frac{2}{c}\partial_{\xi}\partial_{\tau}-\frac{1}{c^{2}}\partial_{\tau}^{~2}\right)
    \left(k_{p}^{~2}\displaystyle\frac{n}{n_{0}}\beta_{z}\right),\\
    \partial_{\xi}^{~2}\phi & = & k_{p}^{~2}\displaystyle\frac{\delta n}{n_{0}}-6\kappa (4\pi e)^{2}\left(\delta n\phi-\frac{n}{\gamma}a^{2}\right)\delta
    n-k_{p}^{~2}\sigma\left(\frac{2}{c}\partial_{\xi}\partial_{\tau}-\frac{1}{c^{2}}\partial_{\tau}^{~2}\right) \displaystyle\frac{\delta n}{n_{0}},
    \label{eqs:8}
  \end{array}\nonumber\\
\end{eqnarray}
and
\begin{eqnarray}
    \frac{1}{c}\partial_{\tau}n & = & \partial_{\xi}\left[n(1-\beta_{z})\right], \label{eqs:9-1}\\
    \partial_{\xi}\left[\gamma(1-\beta_{z})-\phi\right] & = & -\frac{1}{c}\partial_{\tau}(\gamma\beta_{z}). \label{eqs:9-2}
\end{eqnarray}

Assuming $\omega \gg \omega_{p}$, we can use the quasistatic approximation, $\left|\displaystyle\frac{1}{c}\partial_{\tau}\int_{\xi}^{0}nd\xi'\right| \ll n_{0},~
\left|\displaystyle\frac{1}{c}\partial_{\tau}\int_{\xi}^{0}(\gamma\beta_{z})d\xi'\right| \ll 1 $, and obtain the plasma wave hydrodynamic equations,
\begin{eqnarray}
    \frac{\delta n}{n_{0}} & = & \frac{1}{2}\left[\frac{1+a^{2}}{(1+\phi)^{2}}-1\right], \label{eqs:10-2}\\
    \beta_{z} & = & \frac{1+a^{2}-(1+\phi)^{2}}{1+a^{2}+(1+\phi)^{2}}, \label{eqs:10-2}\\
    \gamma & = & \frac{1+a^{2}+(1+\phi)^{2}}{2(1+\phi)}. \label{eqs:10-3}
\end{eqnarray}

Representing the intense laser-plasma interactions with the contribution of the vacuum-polarization effects, the coupling Eq.\eqref{eqs:8} can be simplified as
\begin{equation}
    \displaystyle\bigg(\frac{2}{c}\partial_{\xi}\partial_{\tau}-\frac{1}{c^{2}}
    \partial_{\tau}^{~2}\bigg)a = k_{p}^{~2} \tilde{\eta}\frac{a}{1+\phi},
    \label{eqs:11-1}
\end{equation}
\begin{equation}
    \partial_{\xi}^{~2}\phi =  \displaystyle\frac{k_{p}^{~2}}{2}
    \tilde{\eta}\bigg[\frac{1+a^{2}}{(1+\phi)^{2}}-1\bigg],
    \label{eqs:11-2}
\end{equation}
and
\begin{eqnarray}
    \tilde{\eta} & = & 1-\Delta_{HE}-\Delta_{D} \nonumber\\
    & = & \displaystyle 1-\Lambda^{3}n_{0}
    \bigg[\frac{(a^{2}+\phi^{2})(2+\phi)}{(1+\phi)^{2}}+\frac{4\pi}{k_{p}^{~2}}
    (\frac{2}{c}\partial_{\xi}\partial_{\tau}-\frac{1}{c^{2}}\partial_{\tau}^{~2})\bigg],
    \label{eqs:11-3}
\end{eqnarray}
where $\tilde{\eta}$ is the correction factor, $\Lambda^{3}=({4}/{15})\alpha^{2} \lambdabar_{c}^{~3}$ is the coupling coefficient of vacuum-polarization , and $\lambdabar_{c}={\hbar}/{m_{e}c}$ is the reduced Compton wavelength. Eqs.\eqref{eqs:11-1} and \eqref{eqs:11-2} are the expressions of the 1D nonlinear coupling fields containing the contribution of the vacuum-polarization effects. Eq.\eqref{eqs:11-1} mainly formulates the propagation characteristics of extreme laser through the plasma, and Eq.\eqref{eqs:11-2} is about the evolution of plasma wave induced by the extreme laser beam. The first term on the right of Eq.\eqref{eqs:11-3} is the classical one, the second term $\Delta_{HE}$ is attributed to the H-E correction and the third term $\Delta_{D}$ to the derivative correction. Without considering the contribution of vacuum-polarization effects, i.e., $\Delta_{HE} \sim \Delta_{D} \rightarrow 0, \tilde{\eta} \sim 1$, the Eqs.\eqref{eqs:11-1} and \eqref{eqs:11-1} reduced to the nonlinear classical equations founded by Esarsy et al.\cite{7,13}. This coupling model with the vacuum-polarization effects can help us to investigate the transmission characteristics of laser-plasma interaction under the extreme field laser.

\section{Conclusions and Discussion}

In this work, a 1D nonlinear model of laser-plasma interaction containing vacuum-polarization corrections has been studied. This nonlinear model of the coupling field describes the impact of extreme laser on plasmas. We will focus the attention on how to significantly enhance the effects of vacuum polarization in the coupling processes as follows.

Taking $a \gg 1$ into account, the H-E correction rate, $\Delta_{HE} = \Lambda^{3}n_{0} {(a^{2}+\phi^{2})(2 +\phi)}/{(1+\phi)^{2}}$, can be simplified as $\Lambda^{3}n_{0}a^{2}$. According to the aforementioned relationship of $a$ and $I$, it is easy to obtain the relation as $n_{0}a^{2} \sim 816({\omega_{p}}/{\omega})^{2}I$. Assuming ${\omega_{p}}/{\omega} \sim 0.1$ and $I \sim 10^{25}W/cm^{2}$, the H-E correction rate $\Delta_{HE}$ is closed to $10^{-11}$.

The derivative correction rate, $\Delta_{D}=\Lambda^{3}n_{0}({4\pi}/{k_{p}^{~2}})(2c^{-1}\partial_{\xi}\partial_{\tau}-c^{-2}\partial_{\tau}^{~2})$, is reduced to $\Lambda^{3}4\pi\chi({\omega}/{\omega_{p}})^{2}n_{0}$ under the quasi-plane-wave approximation, where $\chi=1-{k^{2}c^{2}}/{\omega^{2}}$ is the magnetic susceptibility of plasma determined by specific systems. Assuming that ${\omega_{p}}/{\omega} \sim 0.01, \chi \sim 1$, and $n_{0}\sim 10^{25}cm^{3}$, the derivative correction rate is $\Delta_{D} \sim 10^{-6}$.

\begin{figure}[!hbt]
\includegraphics[width=0.70\textwidth]{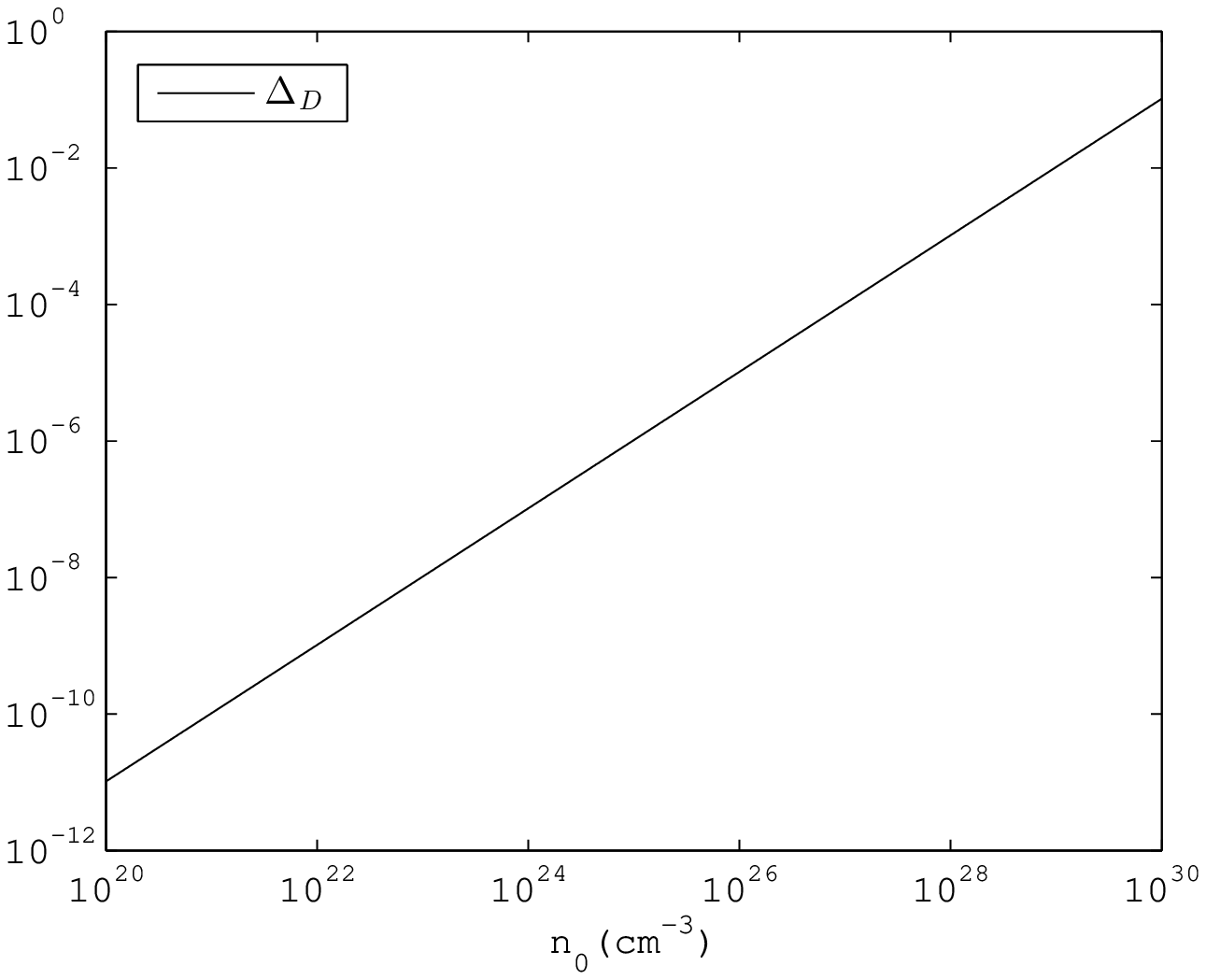}
\caption{~The magnitude of derivative correction rate $\Delta_{D}$ in different plasma density $n_{0}$ for $\chi \sim 1$, $I = 10^{-25}cm^{-3}$ and ${\omega_{p}}/{\omega}=0.01$.}.\label{P1}
\\
\includegraphics[width=0.70\textwidth]{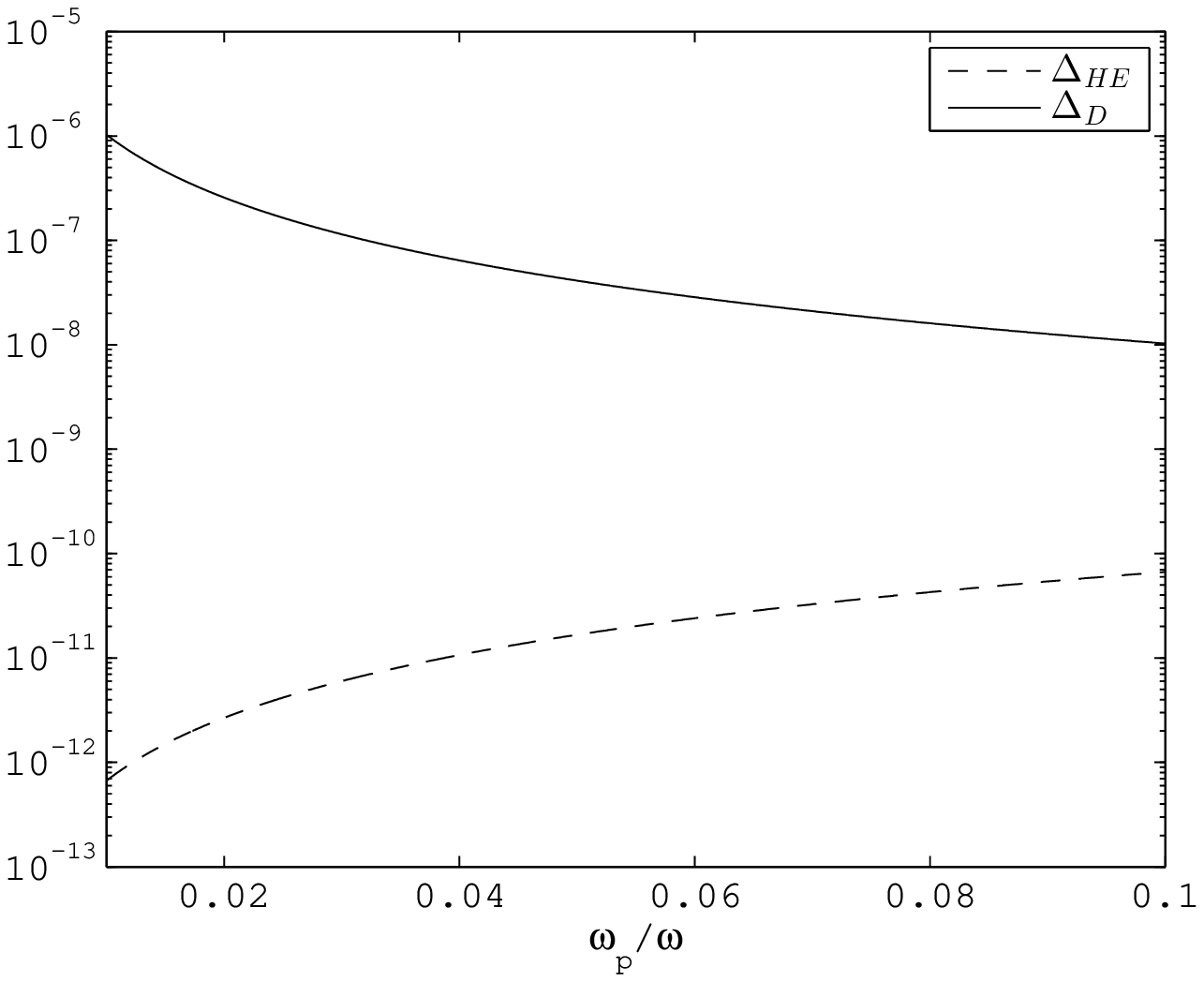}
\caption{~The magnitude of vacuum-polarization correction rates $\Delta_{HE}$ and $\Delta_{D}$ in different frequency factor ${\omega_{p}}/{\omega}$ for $\chi \sim 1$, $I = 10^{-25}W/cm^{2}$ and $n_{0}=10^{25}cm^{-3}$.}.\label{P2}
\end{figure}

The H-E correction of vacuum-polarization is decided by the laser intensity directly. It is calculated that the H-E correction rate $\Delta_{HE}$ is no more than $10^{-10}$
when laser beams with $I\sim10^{25}W/cm^{2}$ travel in an underdense plasma. In a dense plasma, the H-E correction is weakened but more complicated and worthy for further studying,
whereas the derivative effect becomes obvious. Especially by using driving laser beams with ultra-high frequency, the observation of the vacuum-polarization effects is expected to be implemented. Figs. \ref{P1} and \ref{P2} show the relationship between the rates of the vacuum-polarization corrections and the important parameters in the laser-plasma interaction.

In future ICF experiments, which are laser-plasma interactions with the highest plasma density in the experimental environment, electron density $n_{0}$ approaching $10^{-26}cm^{-3}$ is expected to become possible. For this extremely dense states, the derivative correction rate is $\Delta_{D} \sim 10^{-5}$. In astrophysics, with $n_{0} \sim 10^{-30}cm^{-3}$ in the compact objects such as dwarfs\cite{18}, the magnitude of $\Delta_{D}$ can reach $10^{-1}$ further.

\begin{acknowledgments}
This work was partly supported by the Shanghai Leading Academic Discipline Project (Project No. S30105) and the Shanghai Research Foundation (Grant No. 07dz22020).
\end{acknowledgments}

\end{document}